\documentclass[sigconf, 10pt]{acmart}
\renewcommand\footnotetextcopyrightpermission[1]{} 

\usepackage{booktabs} 

\usepackage{amsmath,amssymb,amsfonts}
\usepackage{algorithmic}
\usepackage{graphicx}
\usepackage{textcomp}
\usepackage{xcolor}
\usepackage{tikz}
\usetikzlibrary{patterns}
\def\BibTeX{{\rm B\kern-.05em{\sc i\kern-.025em b}\kern-.08em
    T\kern-.1667em\lower.7ex\hbox{E}\kern-.125emX}}

\newtheorem{definition}{Definition}    
\newtheorem{observation}{Observation}    

\usepackage{algorithm}
\usepackage{multirow}
\usepackage{booktabs}
\usepackage{dcolumn}
\usepackage{rotating} 
\usepackage{arydshln}
\usepackage{listings}
\usepackage{amssymb}
\usepackage{wrapfig}
\usepackage{balance}
\usepackage{url}
\usepackage{hhline}
\usepackage{pgfplots}
\usepackage{adjustbox}
\urldef{\mailsa}\path||    
\usepackage{color}
\definecolor{mygreen}{rgb}{0,0.6,0}
\definecolor{mygray}{rgb}{0.5,0.5,0.5}
\definecolor{mymauve}{rgb}{0.58,0,0.82}
\newcommand{\deepsee}{DeepSolver}

\begin{document}

\title{Constraint Solving with Deep Learning \\for Symbolic Execution}

%
%

\author{Junye Wen$^1$, Mujahid Khan$^1$, Meiru Che$^2$, Yan Yan$^1$, Guowei Yang$^1$}
\affiliation{~}
\affiliation{$^1$Department of Computer Science, Texas State University, San Marcos, TX 78666, USA}
\email{junye.wen@txstate.edu, mujahid.khan@txstate.edu, tom_yan@txstate.edu, gyang@txstate.edu}
\affiliation{$^2$Department of Computer Science, Concordia University Texas, Austin, TX 78726, USA}
\email{meiru.che@concordia.edu}




\begin{abstract}
Symbolic execution is a powerful systematic software analysis technique, but suffers from the high cost of constraint solving, which is the key supporting technology that affects the effectiveness of symbolic execution. Techniques like Green and GreenTrie reuse constraint solutions to speed up constraint solving for symbolic execution; however, these reuse techniques require syntactic/semantic equivalence or implication relationship between constraints. This paper introduces \deepsee, a novel approach to constraint solving with deep learning for symbolic execution. Our key insight is to utilize the collective knowledge of a set of constraint solutions to train a deep neural network, which is then used to classify path conditions for their satisfiability during symbolic execution. Experimental evaluation shows \deepsee\ is highly accurate in classifying path conditions, is more efficient than state-of-the-art constraint solving and constraint solution reuse techniques, and can well support symbolic execution tasks.
\end{abstract}

\maketitle

\section{Introduction} \label{introduction}

Forward symbolic
execution~\cite{King:76,Clarke:76,GodefroidKS05,SenA06,Pasareanu:2010:SPS:1858996.1859035,CadarDE08}
is a powerful technique for systematic exploration of program
behaviors, and provides a basis for various software testing and
verification techniques, such as program equivalence checking,
regression analysis, and continuous
testing~\cite{SiegelMAC08,WhalenGMPTV10,Person:PLDI2011}.  Symbolic
execution executes a program with symbolic values instead of concrete
values, and enumerates the program paths up to a given bound. For each
path it explores, symbolic execution builds a path condition, i.e.,
constraints on the symbolic inputs to follow the corresponding
path. During symbolic execution, off-the-shelf constraint
solvers~\cite{Barrett:2007:CVC:1770351.1770397,Z3Solver} are
used to check the satisfiability of path conditions whenever they
are updated. If a path condition becomes unsatisfiable, the
corresponding path becomes infeasible and is discarded in symbolic
execution.

As the most time-consuming task in symbolic execution, constraint
solving is the key supporting technology that affects the
effectiveness of symbolic execution. The advances in constraint
solving techniques, for example, by leveraging multiple decision
procedures in synergy~\cite{Z3}, have enabled symbolic execution to be
applicable to larger programs. However, despite these technological
advances, symbolic execution still suffers from the high cost of
constraint solving. Several techniques have been developed to speed up
constraint solving for symbolic execution by reusing previous solving
results~\cite{Yang:2013:MTM:2486788.2487001,Visser:2012:GRR:2393596.2393665,Jia:2015,Makhdoom:2014:ISE:2642937.2642961,6825678}.
Various forms of results caching are utilized, so that solutions of
path conditions encountered in previous analysis can be reused without
calling a constraint solver. As a result, the total number of solver
calls as well as the corresponding time cost is reduced.  For example,
Green~\cite{Visser:2012:GRR:2393596.2393665} uses an in-memory
database Redis~\cite{Redis} to store path conditions and their
constraint solutions as key-value pairs, in which key is a path
condition string and value is a Boolean value showing whether the
corresponding path condition is satisfiable or not, and reuses
constraint solutions based on string
matching. GreenTrie~\cite{Jia:2015} further improves the reuse rate of
previous constraint solutions by applying logical reduction and
logical subset and superset querying for given constraints. However,
such reuse techniques require syntactic/semantic equivalence or
implication relationship between constraints. If the equivalence or
implication relationship is not satisfied, these reuse techniques are
able to reuse previous constraint solutions.

In this paper, we introduce \deepsee, a novel approach to constraint
solving with deep learning for symbolic execution. Our key insight is
to utilize the collective knowledge of a set of constraint solutions
to train a deep neural network, which is then used to classify path
conditions for their satisfiability during symbolic execution.  
Deep learning is a popular technique which has found many applications
recently~\cite{Krizhevsky:2017:ICD:3098997.3065386,socher-etal-2013-recursive,5704567,Tolstikhin2018WassersteinA}. It
uses an existing dataset to train a system, similar to the learning
process of a biological neural network, so that the system can process
complex data inputs without being programmed in detail with the
task-specific rules. It has been proved to be effective and efficient
for difficult classification problems such as image
recognition\cite{DBLP:journals/corr/HeZRS15}.  Rather than reusing
each individual constraint solution, \deepsee\ uses the whole set of
constraint solutions to train a deep neural network, and then uses the
deep neural network to classify newly encountered path conditions as
``satisfiable'' or ``unsatisfiable''. Thus, \deepsee\ can classify
path conditions during symbolic execution without calling a constraint
solver which is potentially expensive.

%


Using nine Java programs that have all previously been studied in the
symbolic execution literature, we evaluate \deepsee's accuracy and
efficiency in classifying path conditions, compared to Z3 and
GreenTrie, the state-of-the-art constraint solving and constraint
solution reuse techniques. We also evaluate how \deepsee\ supports
symbolic execution compared to GreenTrie.

We make the following contributions in this paper:

\begin{itemize}
\item We introduce the idea of constraint solving with deep learning
  for symbolic execution. To the best of our knowledge, this is the
  first work on using deep learning for speeding up constraint solving
  in symbolic execution.
\item We design an algorithm for vectorizing a path condition to a
  matrix that enables training of deep neural networks for classifying
  path conditions in symbolic execution.  
\item We design an algorithm for symbolic execution with \deepsee\ for
  constraint solving, which addresses the misclassification errors
  introduced by deep learning and makes \deepsee\ useful in practice.
\item We present an experimental evaluation of \deepsee\ on nine Java
  subjects, which shows that \deepsee\ is highly accurate in
  classifying path conditions for their satisfiability, is more
  efficient than state-of-the-art constraint solving and constraint
  solution reuse techniques, and can well support symbolic execution
  tasks.
\end{itemize}

\section{Background} \label{background}
This section introduces the background on symbolic execution and deep
neural networks.

\subsection{Symbolic Execution} \label{symbolicExe} 

Symbolic execution~\cite{King:76,Clarke:76,Pasareanu:2010:SPS:1858996.1859035,CadarDE08} is a powerful, systematic program analysis
technique. In contrast to concrete execution which takes concrete
values as input and executes only one program path, symbolic execution
executes a program with symbolic values and systematically explores
all program paths up to a given
bound. 
For each path it explores, symbolic execution builds a path condition
($PC$), i.e., constraints on the symbolic inputs to follow the
corresponding path.  
During symbolic execution, off-the-shelf constraint
solvers~\cite{Barrett:2007:CVC:1770351.1770397,Z3Solver} are used to check the the satisfiability of path
conditions whenever they are updated. If a path condition becomes
unsatisfiable, the corresponding path becomes infeasible and is
discarded in symbolic execution.  The state of a symbolically executed
program includes the (symbolic) values of program variables and a
$PC$. A symbolic execution tree characterizes all execution paths
explored during symbolic execution. Each node represents a symbolic
program state, and each arc represents a transition between two
states.


\begin{figure}[t]
\begin{lstlisting}
int example(int x, int y){
  if (x > y){ 
    if (y > 0) 
        x = y + x;
    else
        x = y - x;	
  }else{
    if (x > 0) 
        y = x + y;
    else
        y = x - y;	
  }
}
\end{lstlisting}
\caption{Example program.}
\label{fig:program}
\end{figure}

\begin{figure}[!ht]
\centerline{\includegraphics[width=0.50\textwidth]{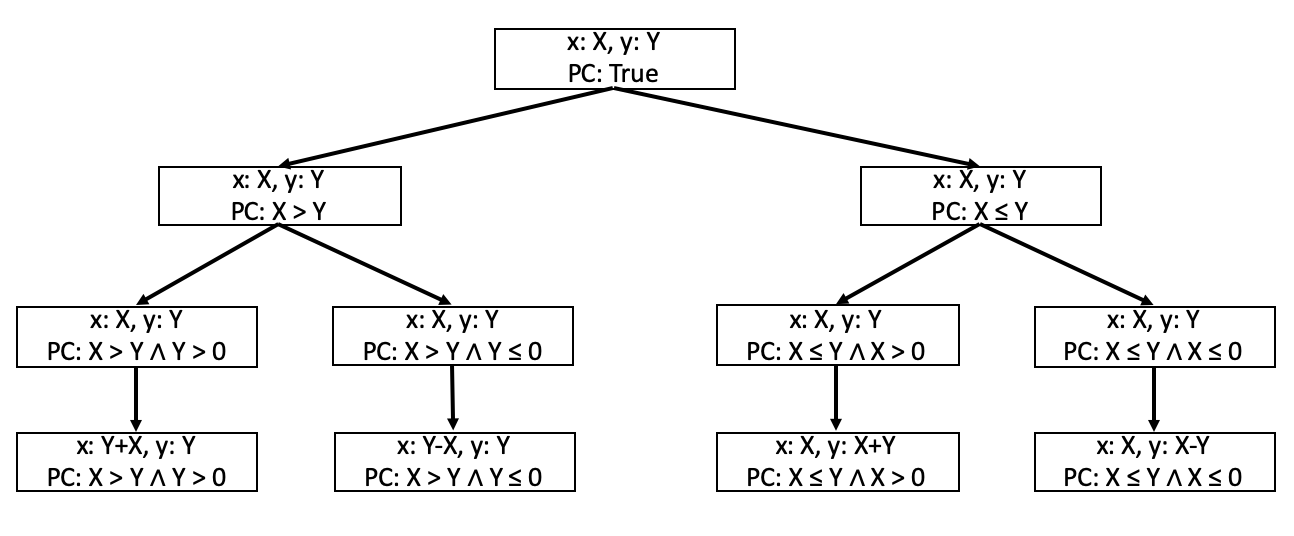}}
\vspace*{-0.1in}
\caption{\label{exampleTree}Symbolic execution tree for the example program}
\vspace*{-0.1in}
\end{figure}

We illustrate symbolic execution on a simple example program in Figure
\ref{fig:program}, which has two integer inputs: {\tt x} and {\tt y}.
For this example, symbolic execution explores four feasible paths
shown in the symbolic execution tree in Figure
\ref{exampleTree}. Initially, $PC$ is {\em True} and {\tt x}, {\tt y}
have symbolic values $X$ and $Y$, respectively. At each branch point,
all {\em choices} are examined with assumptions about the inputs to
choose between alternative paths, while $PC$ is updated
accordingly. For example, after the execution of statement 2, both
{\tt then} and {\tt else} alternatives of the {\tt if} statement are
checked, and $PC$ is updated with different conditions as the
condition is met or violated. Whenever $PC$ is updated, a constraint
solver~\cite{Barrett:2007:CVC:1770351.1770397,Choco,Z3Solver} is called to check its satisfiability. When
$X > Y \wedge Y > 0$ evaluates to $true$ at line $3$ in the source
code, the expression $Y + X$ is computed and stored as the value of x;
when $X > Y \wedge !(Y > 0)$ evaluates to $true$ at line $3$ in the
source code, the expression $Y - X$ is computed and stored as the
value of x.


\begin{figure*}[htb]
\begin{center}
\includegraphics[scale= 0.4]{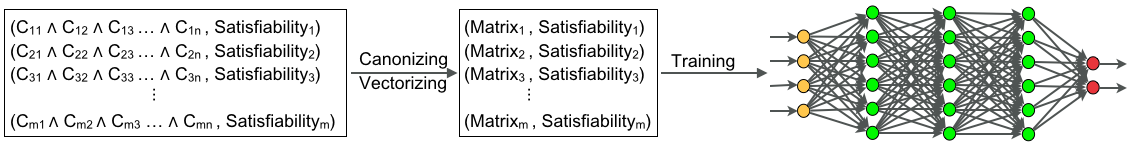}
\caption{Training a DNN with existing constraint solutions}
\label{fig:trainingProcess}
\end{center}
\end{figure*}

Symbolic execution is a widely used technique for different software
analysis purposes such as generating test cases, automatically
checking programs against annotated properties, and detecting
infeasible paths in a program~\cite{GodefroidKS05,10.1007/978-3-642-23702-7_12,Pasareanu:ISSTA08,SenA06,Yang:2012:MSE:2338965.2336771}. However, it suffers from the high cost
of constraint solving, which is the key supporting technology that
affects the effectiveness of symbolic execution.
$PC$ accumulates the constraints on the inputs in order for an
execution to follow the particular associated path, and becomes more
and more complex as the path goes deeper in the symbolic execution
tree. The complexity of $PC$ increases when more constraints are
accumulated, non-linear calculation are performed, or more symbolic
variables are involved. The more complex a $PC$ becomes, the more
difficult it is for a constraint solver to check its
satisfiability. Despite the recent advances in constraint
solving~\cite{SouzaBdP11,CDW14,SMTCOMP2019} which have enabled
symbolic execution to be applicable to larger programs, constraint
solving remains a bottleneck of symbolic execution.

\subsection{Deep Neural Networks} \label{deeplearning}

Deep neural networks (DNNs) have been widely used in many artificial
intelligence areas, such as computer
vision~\cite{Krizhevsky:2017:ICD:3098997.3065386}, natural language
processing~\cite{socher-etal-2013-recursive}, and speech
recognition~\cite{5704567}. In a deep learning model, many layers of
information processing stages in hierarchical architectures are
utilized for pattern classifications or feature learning
purposes. DNNs use multiple layers to progressively extract higher
level features from raw input.

One common usage of DNNs is as {\em classifiers}. Each input data to
the DNN is assigned a pre-set label or class as an output. Each layer
of a DNN is comprised of nodes, termed {\em neurons}, and the nodes
refines and extracts information based on value sent from the previous
layer, and then applies their own function to compute a value for the
next layer.  A typical DNN has one input layer which takes in the
input data, one output layer which generates the final classification
results, and several hidden layers to perform intermediate processing
(e.g., feature extraction). Each neuron computes its output by
applying an {\em activation function} (e.g., ReLu or sigmoid) to the
weighted sum of its inputs according to a unique weight vector and a
bias value.

\section{DeepSolver} \label{DeepSolver}

In this section, we present \deepsee, which consists of two stages:
the first stage trains a DNN using existing constraint solutions (Section~\ref{trainingstage}), and
the second stage uses the trained DNN to classify path conditions for
their satisfiability (Section~\ref{classificationstage}).

\subsection{Training a DNN with Constraint Solutions} \label{trainingstage}


Figure~\ref{fig:trainingProcess} shows the overall process of training
a DNN with existing constraint solutions in the form of {\em
  PC-satisfiability} pair where {\em PC} is the path condition and
{\em satisfiability} is a Boolean value (i.e., {\em True} or {\em
  False}) indicating whether the PC is satisfiable or not.  As DNNs
require the input data to be in the form of a matrix. Thus, we first
canonize and vectorize the PCs to matrices, and then use the matrices
and satisfiability information to train a DNN. We currently only
support linear integer arithmetic path conditions, and will support
other types of path conditions in future work.

\subsubsection{Canonizing} \label{canonizing}

Path conditions generated during symbolic execution do not have a
common pattern by default. For instance, the name space of symbolic
variables differs from subject to subject. {\em Canonizing} transforms
a path condition into a unified
format. 
Each constraint in the path condition is transformed
into a normal form for linear integer arithmetic path conditions,
specifically
$$ax + by + cz + ... + k\, op\, 0,\; where\, op \in \{ =, \neq,
\leqslant \}$$
Other operators including $>$, $<$ and $\geqslant$ are transformed
into the corresponding canonical forms with operators $=, \neq,$ and
$\leqslant$. Meanwhile, the constraints are sorted in a lexicographic
order and then symbolic variables are renamed based on their
appearances in the path condition in left-to-right order. 
For instance, both ${PC}_1: x + y < z \wedge x = z \wedge x - 10 > y$
and ${PC}_2: a + b < c \wedge a = c \wedge a - b > 10$ will be
canonized into a same shape as
$v_0 + v_1 - v_2 + 1 \leqslant 0 \wedge v_0 - v_2 = 0 \wedge -v_0 +
v_1 - 9 \leqslant 0$.
A unified name space of variables can help us vectorize path
conditions into matrices, and eliminate the equivalent records in the
training data set.

Since {\em canonizing} has been used in previous constraint solution
reuse techniques~\cite{Visser:2012:GRR:2393596.2393665, Jia:2015},
this paper only briefly discusses {\em vectorizing}. Please refer
to~\cite{Visser:2012:GRR:2393596.2393665} for more details.

\subsubsection{Vectorizing} \label{vectorizing} 

After {\em canonizing}, all path conditions have the same variable
name space and are in a unified normal form. 
We then perform {\em vectorizing} to turn a path condition in plain
text into a 2-dimensional matrix.

We have two important observations of path condition generated in
symbolic execution:

\begin{observation}
  A path condition is joined by multiple constraints. It is a
  typical conjunctive normal form (CNF) as it uses only $AND$ logical
  operator between constraints.
\end{observation}

\begin{observation}
  In symbolic execution, all variables in a path condition are
  expressed by the symbolic input variables. Thus, each
  constraint can be represented as a multinomial formula on the
  symbolic input variables.
\end{observation}

To further explain our technique, we give the following definitions,
which will be used in the rest of this paper:

\begin{definition}
  A path condition has a dimension of $(d, n, t)$, or it is a
  $(d, n, t)$ path condition, where $d$ is the number of constraints,
  $n$ is the highest degree of a term among all constraints, and $t$
  is the number of symbolic variables in the path condition.
\end{definition}


\begin{definition}
  A $(d, n, t)$ path condition 
  is $linear$ if $n = 1$; otherwise, it is $nonlinear$.
\end{definition}


For instance, $x + y - z + 1 \leqslant 0 \wedge  x - z = 0 \wedge -x + y - 9 \leqslant 0$ has a dimension of (3, 1, 3) and it is a linear path condition. $v{_0^2} + v_0 \times v_1 + v{_1^3} + 1 > 0$ is a nonlinear (1, 3, 2) path condition. 

After canonizing, a constraint is transformed into the form
$c_0 \times v_0 + c_1 \times v_1 + c_2 \times v_2 + ... + k\, op\, 0$,
where $c_n$ is the coefficient of variable $v_n$, $k$ is the constant
term, and $op \in \{ =, \neq, \leqslant\}$. Since the name space of
symbolic variables is unified, a (d, n, t) path condition joined by
constraints in this format can be easily transformed into a
2-Dimensional matrix: each row of the matrix stands for a constraint,
and the columns stand for the coefficient for one symbolic variable,
the constant term, and an integer value used to represent $op$.  The
size of the 2-Dimensional matrix is determined as follows:

The number of rows $X$ is determined by the number of constraints in a
path condition, each row representing one constraint.  Since symbolic
execution generates path conditions with different number of
constraints, we can have two strategies:
we can group path conditions based on the number of their constraints,
and train a DNN for each group; or we can use $padding$ to expand path
conditions that have smaller number of constraints, until they have
the same number of constraints with the path condition that has the
largest number of constraints.
Being a typical CNF, a path condition can be joined by any number of
$true$ conditions without changing its satisfiability. In our
2-Dimensional matrix model, we can simply add rows with all columns
set to $0$. By our design, such a row represents formula $0 = 0$,
which is identically {\em True}. Theoretically, any logic $True$
formula can be used as padding. However, 
it may potentially impact the accuracy of the trained neural
network. 
Thus, in this paper, we choose to use the first strategy and train
multiple DNNs to handle path condition with different number of
constraints. We leave the second strategy that trains one single
neural network for all path conditions with different sizes for our
future work.

\begin{figure*}[t!]
\begin{center}
\includegraphics[scale=0.45]{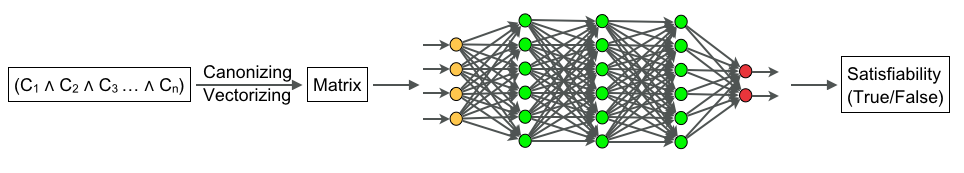}
\caption{Classifying a Path Condition Using a DNN}
\label{fig:applyingProcess}
\end{center}
\end{figure*}

For a linear  path condition with $t$ symbolic variables, the number of columns $Y  = t + 2$,
where $t$ means that we need $t$ columns to represent the coefficients
of $t$ symbolic variables, while the constant value $2$ means that we
need $2$ extra columns for the constant term and an integer
representing the operator. In our model, since we only have three
different operators, we assign value $0$ for $=$, $1$ for $\neq$, and
$2$ for $\leqslant$, respectively. 

  \begin{algorithm}[htb]
\caption{Algorithm for vectorizing canonized linear PC to a corresponding matrix} 
\label{algorithmVectorization} 
\begin{algorithmic}[1]
\footnotesize

\REQUIRE Canonized path condition $PC$, which is linear and
in shape of $BC_0 \wedge BC_1 \wedge ... \wedge BC_m$, where $BC_m$ is
in shape of
$c_0 \times v_0 + c_1 \times v_1 + ...  + c_n \times v_n + k\, op\, 0$
($op \in \{ =, \neq, \leqslant \}$)

\STATE $X$ $\gets$ $m$+1;
\STATE $Y$ $\gets$ $n$+3;
\STATE Array[$X$][$Y$] $Matrix$ $\gets$ empty;

\STATE List $BCS$ $\gets$ $PC$ split by $''\wedge''$;
\STATE $i$ $\gets$ 0;
\WHILE{$i < X$}
	\IF {$op$ in $BCS[i]$ is $''=''$}
		\STATE $Matrix[i][Y-1]$ $\gets$ 0;
	\ELSIF {$op$ in $BCS[i]$ is $''\neq''$}
		\STATE $Matrix[i][Y-1]$ $\gets$ 1;
	\ELSIF {$op$ in $BCS[i]$ is $''\leqslant''$}
		\STATE $Matrix[i][Y-1]$ $\gets$ 2;
	\ENDIF
	\STATE $BCS[i]$ $\gets$ $BCS[i]$ remove $op$;
	\STATE List $Terms$ $\gets$ $BCS[i]$ split by $''+''$;
	
	\FOR {\textbf{all} $term$ in $Terms$}
		\IF {$term$ in shape of  $c_j \times v_j$}
			\STATE $Matrix[i][j]$ $\gets$ $c_j$;
		\ELSE 
			\STATE \COMMENT{$term$ is the constant term $k$}
			\STATE $Matrix[i][Y-2]$ $\gets$ $k$;
		\ENDIF
	\ENDFOR
\ENDWHILE
\RETURN $Matrix$;

\end{algorithmic}
\end{algorithm}
  Algorithm
  \ref{algorithmVectorization} shows how to vectorize a linear path
  condition into a matrix, after the path condition has been canonized. 
  We first initialize the $Matrix$ by the number of
  constraints and the largest index of symbolic variable in the
  path condition (Lines $1-3$). The path condition is first split by
  ``$\wedge$'' into a list of constraints $BCS$ (Line
  $4$). Each constraint in $BCS$ is checked to set up a row in
  $Matrix$ (Lines $5-24$). For each constraint, we first check
  its operator and set the corresponding item in the row as $0$, $1$
  or $2$ (Lines $7-13$). Then the constraint is further broken
  down to a list of terms $Terms$ by ``$+$'' after removing the
  equation operator $op$ (Lines $14-15$). As we go through each term
  in the $Terms$, if the term is in a shape of $c_j \times v_j$, we
  set the $j$-th item in the row as $c_j$ (Lines $17-18$); otherwise,
  the term is a constant value $k$, which is used as the value of the
  second last item in the row (Lines $20-21$). After all 
  constraints are processed, we return $Matrix$ as the final
  vectorized result of the path condition (Line $25$).

  With this algorithm, any path condition can be transformed into a
  2-Dimensional matrix and expanded to a larger equivalent matrix if
  needed. For instance, the previous example path condition
  $x + y - z + 1 \leqslant 0 \wedge x - z = 0 \wedge -x + y - 9
  \leqslant 0$
  can be transformed in its original (3, 1, 3) path condition matrix
  format in size $3 \times 5$ (we assign the second last column for
  constant term, and last column for the operator) as:
$$
\left[
\begin{array}{ccccc}
1	& 1	& -1	& 1	& 2		\\
1	& 0	& -1	& 0	& 0		\\
-1	& 1	& 0	& -9	& 2
\end{array}
\right]
$$

Also, if needed (e.g. when another path condition with the same number
of constraints but more symbolic variables), the algorithm can
be easily modified to expand the matrix to a larger yet equivalent
matrix in size $3 \times 6$ as:
$$
\left[
\begin{array}{cccccc}
1	& 1	& -1	& 0	& 1	& 2		\\
1	& 0	& -1	& 0	& 0	& 0		\\
-1	& 1	& 0	& 0	& -9	& 2		
\end{array}
\right]
$$


\subsection{Classifying Path Conditions Using a DNN} \label{classificationstage}

Figure~\ref{fig:applyingProcess} shows the steps involved to classify
a path condition generated in symbolic execution using a DNN that have
has been trained with existing constraint solutions. 
The path condition also goes through the same $canonizing$
and $vectorizing$ as in the training stage, in order to get its
corresponding matrix. The vectorized path condition in form of a
matrix is then sent to a previously trained DNN based on its size
(defined by the X and Y dimensions of the matrix) and the
classification output (satisfiability of path condition) is then
returned to symbolic execution to decide whether the corresponding
path is feasible or not.

After training, we only require the DNNs and their corresponding
vectorization algorithms to classify a path
condition. 
As long as a path condition can be transformed into a matrix that is
acceptable by one of the previously-trained DNNs, our approach is
capable to classify the path condition for its satisfiability.
Moreover, since DNNs are trained off-line, i.e. they are totally
separated from symbolic execution runs, users can train a different
DNN while the classification with current DNNs is still in progress.
This ensures that our framework can be updated and expanded with
minimum extra work to check more complicated path conditions.

\section{Symbolic Execution with \deepsee} \label{SEwithDS}


Ideally, a DNN should be about to reach $100\%$ accuracy in
classification. However, in practice this goal is extremely difficult
to achieve, and in most cases $100\%$ accuracy indicates the
possibility of over-fitting problem~\cite{over-fitting}. An
over-fitting problem happens when a classifier is overly refined to a
certain data set and thus cannot be applied on other inputs while
keeping a high accuracy. As a result, DNNs are usually used with a
high accuracy while tolerating potential misclassifications.

When DNNs are used for satisfiability checking of path conditions in
symbolic execution, the misclassification problem can make symbolic
execution unsound, and thus we need to address the problem.  In
particular, there are two types of misclassification errors: a
satisfiable PC is classified as unsatisfiable (Type I
misclassification) or an unsatisfiable PC is classified as satisfiable
(Type II misclassification). We discuss in the following how to deal
with each of the two types of misclassification errors.

\subsection{Type I Misclassification} \label{soundness-fn}

When a satisfiable PC is classified as unsatisfiable, the
corresponding path is incorrectly identified as infeasible. Since
symbolic execution will not continue the exploration of a path when it
becomes infeasible, this type of misclassification causes symbolic
execution to explore fewer states and must be avoided. To address this
problem, we propose to double-check the questionable classification
result when a PC is classified as unsatisfiable by calling a
conventional constraint solver. This extra constraint solving of
course will introduce an overhead. However, this overhead is
relatively small for two reasons. First, in most cases, the number of
infeasible paths explored in symbolic execution is relatively
small compared to the number of feasible paths. 
Second, assuming the DNN models are highly accurate, the chance of a
Type I misclassicaction happening is low. Therefore, we do not have to
frequently double-check the classification result, and thus the
overhead introduced by calling a conventional constraint solver is
small. 

\subsection{Type II Misclassification} \label{soundness-fp} 

On the other hand, when a unsatisfiable PC is classified as
satisfiable, the corresponding path is incorrectly identified as
feasible, and thus symbolic execution may continue exploring states
that are in fact not feasible.  For intermediate states, instead of
double-checking ``unsatisfiable'' classification results to avoid Type
I misclassification errors, we ignore the possible Type II
misclassification errors based on the following two observations:
First, it is safe to explore some infeasible states.  Second, assuming
the classification of our approach is highly accurate, it is very
likely that symbolic execution based on DNNs will explore few such
infeasible states and thus the extra cost is low. Consider an
infeasible path with a condition $PC$ as an example. If the
classification accuracy of our approach is over 90\%, the chance of
Type II misclassification less than 10\%. When such misclassification
happens, symbolic execution will continue on this path and explore
another infeasible path with the updated path condition
$PC' = PC \wedge c$, where $c$ is the new constraint collected along
the path. Assuming $PC$ and $PC'$ are treated independent in DNN
classification, the chance of misclassifying both of them is only
$10\% \times 10\% = 1\%$. Therefore, the chance of continuous
misclassification drops significantly as the exploration goes deeper.
In other words, even if an unsatisfiable PC is classified as
satisfiable, it is very likely that the exploration will only explore
very small number of extra states before a new PC is classified as
unsatisfiable. 

For leaf states, which represent complete paths or paths stopped due
to errors, we call the underlying constraint solver to find input
values to test the corresponding path or to trigger the detected
errors, for the two most popular application of symbolic execution:
test case generation and error detection.

\subsection{Algorithm} \label{deepSeeAlgorithm}
\begin{algorithm}[htb]
\caption{Symbolic Execution with \deepsee}
\label{algorithmDeepSee} 
\begin{algorithmic}[1]
\footnotesize

%
%
%

\REQUIRE Trained DNN model collection $M$
\STATE Test Suit $T$ $\gets$ $\emptyset$;
\STATE $init\_state.PC$ $\gets$ $True$;
\STATE $stack.push(init\_state)$;
\STATE Boolean $\varphi$ $\gets$ $True$;
\WHILE {$\neg stack.empty()$}
	\STATE $s$ $\gets$ $stack.pop()$;
	\STATE $pc$ $\gets$ $s.PC$
	\STATE $\varphi$ $\gets$ check($pc$, $M$);
	\IF {$\varphi$ is $False$ \OR $pc$ is not supported by $M$}
			\STATE $\varphi$ $\gets$ solve($pc$);
	\ENDIF

	\IF {$\varphi$ is $True$}
		\FOR {each instruction $inst$}
			\IF {$inst$ is $if(c)$}
				\STATE \COMMENT{Let $c$ be constraint for $True$ branch}
				\STATE $s'.PC$ $\gets$ $pc \wedge c$;
				\STATE $stack.push(s')$;
				\STATE $s'.PC$ $\gets$ $pc \wedge \neg c$;
				\STATE $stack.push(s')$;
				\STATE \textbf{break};
			\ELSIF {$inst$ is $abort$ or $halt$}
				\STATE Test case $t$ $\gets$ solve($pc$);
				\STATE $T$ $\gets$ $T$ $\cup$ \{$t$\};
				\STATE \textbf{break};
			\ELSE
				\STATE $s$ $\gets$ $execute(inst, s)$;
			\ENDIF
		\ENDFOR
	\ENDIF
\ENDWHILE
\RETURN $T$

\end{algorithmic}
\end{algorithm} 

We show our algorithm of symbolic execution with \deepsee\ in
Algorithm~\ref{algorithmDeepSee}. It is similar to traditional forward
symbolic execution that uses depth-first search to explore all
feasible paths of a program, except for several key steps to address
the aforementioned problems. In particular, instead of calling a
constraint solver to check the satisfiability of a path condition
$pc$, we first use \deepsee\ to check its satisfiability, noted as
$check(pc, M)$ (Line $8$). This represents the process described in
Section~\ref{classificationstage}, where the {\em pc} is canonized and
vectorized to generate the corresponding matrix. If a DNN
corresponding to the matrix size exists in $M$, a Boolean value of
will be returned.  If there is no DNN in $M$ that could handle the
matrix, the underlying constraint solver will be called instead. Also,
if the classification result shows the {\em pc} is not satisfiable, we
double-check it with the constraint solver to avoid Type I
misclassification as stated in Section \ref{soundness-fn} (Lines
$9-11$). In addition, constraint solver is called to generate input
values for paths that are naturally completed or aborted due to errors
(Lines $22-23$).

\section{Evaluation} \label{evaluation}
This section evaluates \deepsee\ on its performance in classifying path conditions as well as in
supporting symbolic execution.
Our evaluation aims to answer the following four research questions:

\begin{itemize}
\item \textbf{RQ1:} How accurate is \deepsee\ in path condition
  classification?
\item \textbf{RQ2:} How efficient is \deepsee\ in path condition
  classification compared to state-of-the-art constraint solving and
  constraint solution reuse techniques?
\item \textbf{RQ3:} How do the DNN structure and the size of the
  training data impact \deepsee's accuracy and efficiency?
\item \textbf{RQ4:} How well does \deepsee\ support symbolic
  execution?

\end{itemize}


\subsection{Implementation and Subjects} \label{implementation} We train our DNNs
with Keras~\cite{chollet2015keras}, which is a high-level deep learning
API written in Python and is capable of running on top of
TensorFlow~\cite{tensorflow2015-whitepaper}. 
We implement $canonizing$ and $vectorizing$ modules, and symbolic
execution with \deepsee\ in Symbolic Pathfinder
({SPF})~\cite{Pasareanu:2010:SPS:1858996.1859035}, a widely used
open-source symbolic execution framework for Java programs.
Since Keras models cannot be directly run with Java framework, we
convert the trained DNNs into TensorFlow's format and run them with
official TensorFlow Java library.



The subjects chosen for our evaluation are widely used as benchmarks
before for evaluating symbolic execution
techniques~\cite{Spain,inkumsah08:improving,Person:PLDI2011,Yang:2013:MTM:2486788.2487001,SouzaBdP11,Bytecode13,iProperty,ourICSE15,
  wise, Jia:2015}.

\textbf{Traffic Anti-Collision Avoidance System (TCAS)} is a system to avoid air collisions. Its code in C together with $41$ mutants are available at SIR repository~\cite{SIR}. We manually converted the code to Java and only used the original version for this case study. 

\textbf{Wheel Brake System (WBS)} is a synchronous reactive component from the automotive domain. This method determines how much braking pressure to apply based on the environment. The Java model is based on a Simulink model derived from the WBS case example found in ARP 4761~\cite{ARP4761,Joshi05:SafeComp}. The Simulink model was translated to C using tools developed at Rockwell Collins and manually translated to Java. 

\textbf{MerArbiter} is a component of the flight software for NASA JPL's Mars Exploration Rovers (MER). 

\textbf{Red-Black Tree Data Structure} is the code of data structure originally from Suns JDK 1.5. 

\textbf{Dijkstra} is a benchmark developed by Jacob Burnim from University of California, Berkeley. It is an algorithm for finding the shortest paths between nodes in a graph, which may represent road networks for instance. 

\textbf{TSP} is a benchmark solution for Traveling Salesman Problem. This subject is developed by Sudeep Juvekar and Jacob Burnim from California, Berkeley. 

\textbf{Rational} is a case study for computing greatest common divisor and its related operations on rational numbers. 

\textbf{BinTree} implements a binary search tree with element insertion, deletion. 


\textbf{BinomialHeap} is a Java implementation of binomial heap.


\subsection{DNN Training} \label{DNN} 

As the structure of a DNN may affect its accuracy and efficiency, in
this evaluation we compare two different structures of DNN based on
the number of hidden layers and the number of neurons in each
layer. One {\sl small} structure has 5 hidden layers and 5 neurons in
each layer (we refer to this size as $5 \times 5$), and one {\sl big}
structure has 10 hidden layers and 10 neurons in each layer (we refer
to this size as $10 \times 10$). Both structures use dense connection
with ReLu activation
function~\cite{DBLP:journals/corr/abs-1803-08375}.

Another important factor in deep learning techniques is the {\em
  training data}. 
Generally speaking, a dataset of thousands of records is enough to
train an applicable DNN. A training dataset for image or video
processing with deep learning usually has thousands of records. For
example, UCF-101~\cite{DBLP:journals/corr/abs-1212-0402} has 13K
videos, and HMDB-51~\cite{6126543} has 6.8K videos.
In our evaluation, we use two different training datasets: A {\sl
  small} dataset and a {\sl large} dataset.
The small dataset consists of constraint solutions from running
symbolic execution with Z3~\cite{Z3Solver} on \texttt{TCAS}, \texttt{WBS} and
\texttt{MerArbiter}. The large dataset consists of all the constraint
solutions in the small dataset plus additional constraint solutions
from running symbolic execution with Z3 on the mutants of the three
subjects.
Meanwhile, we noticed that due to different computation orders used in
different subjects, there are logically equivalent records in the data
sets even after canonization. For instance,
$(((2 \times x) + (3 \times y)) - (4 \times z)) + 1 \leqslant 0$ and
$((2 \times x) + ((3 \times y) - (4 \times z))) + 1 \leqslant 0$ are
treated as two records in the datasets although they are identical
after being vectorized in to the Matrix. We removed all logically
equivalent path conditions from the training datasets, for duplicate
records can interfere with our training process and lead to potential
over-fitting problem. Finally, the small dataset has $514,230$
records, while the large dataset has $1,417,691$ records.

We group path conditions based on the number of constraints involved
in the path conditions, and train a DNN for each group. The
path conditions in our training data have at most 28 constraints
involved. Therefore, the size of matrix used in our evaluation ranges
from $22 \times 11$ to $22 \times 28$, which represent path conditions
with 11-28 constraints and 20 different symbolic variables (20 columns
for the 20 symbolic variables and 1 column for the constant term, and
1 column for the operator mark).  Table~\ref{tab:EvaResult2} shows the
number of records we have for each group of path conditions to train a
DNN. Specifically, we have at least $3,584$ records (using the
small dataset) to train a DNN (for PCs with 11 constraints), and we
have at most $150,212$ records (using the large dataset) to
train a DNN (for PCs with 20 constraints). We did not train DNNs for
path conditions that have $10$ or fewer constraints since the number
of constraint solutions in our dataset for such path conditions is
too small to train DNNs.

\pgfplotsset{ytick style={draw=none}}
\pgfplotsset{xtick style={draw=none}}
\pgfplotsset{every tick label/.append style={font=\large}}
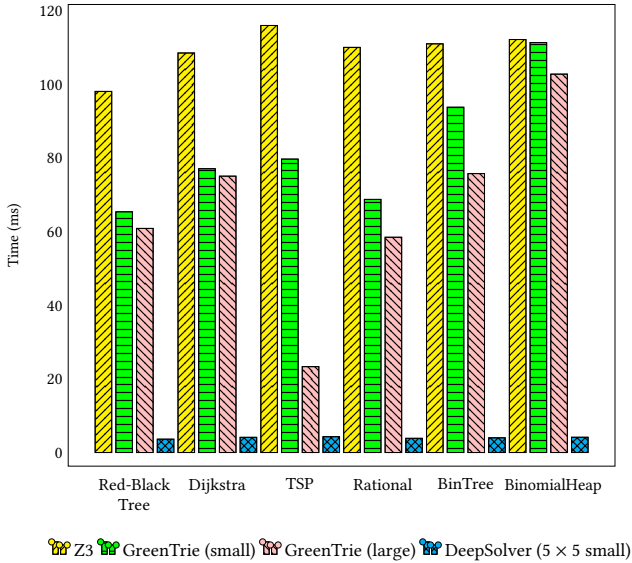
\begin{figure}
\begin{adjustbox}{width=\columnwidth}

\begin{tikzpicture}
\begin{axis}[
	xtick=data,
	width=14cm,
	ylabel=Time (ms),
	enlargelimits=0.05,
	legend style={at={(0.5,-0.15)},
		anchor=north,legend columns=-1},
	ybar=5pt,
	ybar interval=0.8,
	symbolic x coords={Red-Black Tree, Dijkstra, TSP, Rational, BinTree, BinomialHeap, dummy},
    x tick label style={align=center, text width=2cm},
    xlabel near ticks,
    grid=none,
    legend style={nodes={scale=1.2, transform shape}}, 
        legend image post style={mark=*},
        legend style={draw=none}
]
\addplot[black, fill=yellow,postaction={ pattern=north east lines}] coordinates {(Red-Black Tree,98.18) (Dijkstra,108.63) (TSP,116.11) (Rational,110.15) (BinTree,111.12) (BinomialHeap,112.29) (dummy,2)};

\addplot[black, fill=green,postaction={ pattern=horizontal lines}] 
	coordinates {(Red-Black Tree,65.45) (Dijkstra,77.17)
		 (TSP,79.78) (Rational,68.82) (BinTree,93.88) (BinomialHeap,111.47) (dummy,2)};
\addplot[black, fill=pink,postaction={ pattern=north west lines}]
	coordinates {(Red-Black Tree,60.93) (Dijkstra,75.14)
		 (TSP,23.35) (Rational,58.53) (BinTree,75.84) (BinomialHeap,102.88) (dummy,2)};
\addplot [black, fill=cyan,postaction={ pattern=crosshatch}]
	coordinates {(Red-Black Tree,3.63) (Dijkstra,4.13)
		 (TSP,4.30) (Rational,3.86) (BinTree,4.00) (BinomialHeap,4.15) (dummy,2)};

\legend{Z3,GreenTrie (small), GreenTrie (large), DeepSolver (5 $\times$ 5 small)}
\end{axis}

\end{tikzpicture}

\end{adjustbox}
\caption{Comparison of average time cost of satisfiability check}
\label{fig:averCheckingComp}
\end{figure}


\begin{table*}[ht]
  \caption[EvaResult1]{Results of classifying PCs using \deepsee\, compared to  Z3 and GreenTrie.}
\begin{center}
\scalebox{.72}{
\begin{tabular}{|c||c||c||c|c|c|c||c|c|c|c|c|c|c|c|}
\hline

\multirow{4}{*}{Subjects}	& \multirow{4}{*}{\# PCs}		& Z3		& \multicolumn{4}{c||}{GreenTrie}	& \multicolumn{8}{c|}{\deepsee} 	\\ \cline{3-15} 
	&		& \multirow{3}{*}{Time Cost (s)}  & \multicolumn{2}{c|}{\multirow{2}{*}{Time Cost (s)}} & \multicolumn{2}{c||}{\multirow{2}{*}{Reuse Rate}} & \multicolumn{4}{c|}{5X5 DNN}	& \multicolumn{4}{c|}{10X10 DNN}                                    	\\ \cline{8-15} 
							&					&					& \multicolumn{2}{c|}{}	& \multicolumn{2}{c||}{}	& \multicolumn{2}{c|}{Time Cost (s)}	& \multicolumn{2}{c|}{Accuracy}	& \multicolumn{2}{c|}{Time Cost (s)}	& \multicolumn{2}{c|}{Accuracy} \\ \cline{4-15} 
							& 		 			&					& Small			& Large	& Small		& Large 	& Small		& Large	& Small			& Large		& Small		& Large	& Small			& Large		\\ \hline \hline
							

Red-Black Tree	& 1,283		& 125.97		& 83.97		& 78.17		& 28\%		& 34\%		& 4.66		& 4.66		& 98.9\%		& 98.4\%		& 8.57		& 8.94		& 98.5\%		& 98.8\%		\\ \hline
Dijkstra				& 10,582	& 1,149.55	& 816.62		& 795.18		& 32\%		& 57\%		& 43.68	& 44.83	& 97.8\%		& 98.3\%		& 80.47	& 82.77	& 98.2\%		& 99.9\%		\\ \hline
TSP						& 13,195		& 1,532.13	& 1,052.69	& 308.17		& 31\%		& 85\%		& 56.69	& 55.16	& 97.6\%		& 98.3\%		& 105.72	& 107.25	& 99.4\%		& 99.9\%		\\ \hline
Rational				& 716			& 78.86		& 49.27		& 41.91		& 38\%		& 44\%		& 2.76		& 2.84		& 98.7\%		& 99.7\%		& 5.76		& 5.44		& 97.5\%		& 98.0\%		\\ \hline
BinTree					& 3,401		& 377.90		& 319.27		& 257.94		& 35\%		& 42\%		& 13.60	& 13.98	& 97.7\%		& 98.6\%		& 27.21	& 26.45	& 98.5\%		& 99.4\%		\\ \hline
BinomialHeap		& 23,156		& 2,600.19	& 2,581.17	& 2,382.27	& 33\%		& 46\%		& 96.21	& 96.21	& 99.2\%		& 99.5\%		& 189.81	& 184.61	& 97.6\%		& 99.8\%		\\ \hline

\end{tabular}
}
\end{center}
\label{tab:EvaResult1}
\end{table*}

\begin{table}[ht]
  \caption[EvaResult2]{Individual DNN's accuracy for classifying PCs with 11-28 constraints.}
\begin{center}
\scalebox{.65}{
\begin{tabular}{|c|c|c|c|c|c|c|c|}
\hline
{\# Constraints} & \multicolumn{2}{c|}{\# Records in Training Data} & {\#} & \multicolumn{2}{c|}{5X5 DNN}  & \multicolumn{2}{c|}{10X10 DNN}  \\ \cline{2-3} \cline{5-8} 
{in a PC}  & Small & Large & New PCs & Small & Large & Small & Large \\ \hline \hline
11	& 3,896		& 21,397		& 418		& 98.3\%		& 98.5\%		& 98.7\%		& 99.9\% \\ \hline
12	& 7,507		& 29,174		& 634		& 98.2\%		& 99.2\%		& 98.2\%		& 99.2\% \\ \hline
13	& 10,596		& 27,282		& 915		& 98.9\%		& 98.9\%		& 98.3\%		& 98.9\% \\ \hline
14	& 14,684		& 39,405		& 1,429	& 98.7\%		& 98.8\%		& 98.6\%		& 99.0\% \\ \hline
15	& 19,043		& 60,723		& 2,380	& 98.3\%		& 99.0\%		& 98.6\%		& 98.7\% \\ \hline
16	& 25,594		& 86,656		& 3,507	& 99.4\%		& 99.6\%		& 98.0\%		& 99.6\% \\ \hline
17	& 34,119		& 112,415	& 4,547	& 98.1\%		& 98.3\%		& 97.9\%		& 98.8\% \\ \hline
18	& 43,082		& 132,410	& 5,306	& 98.2\%		& 99.8\%		& 98.6\%		& 99.3\% \\ \hline
19	& 51,273		& 145,182	& 5,755	& 98.2\%		& 99.5\%		& 99.3\%		& 99.0\% \\ \hline
20	& 57,062		& 150,212	& 5,512	& 98.5\%		& 98.7\%		& 98.2\%		& 99.7\% \\ \hline
21	& 57,242		& 146,252	& 5,318	& 97.9\%		& 98.4\%		& 99.5\%		& 99.9\% \\ \hline
22	& 53,292		& 131,509	& 4,927	& 99.1\%		& 99.0\%		& 99.5\%		& 99.9\% \\ \hline
23	& 46,396		& 112,405	& 4,139	& 98.6\%		& 98.6\%		& 99.3\%		& 99.6\% \\ \hline
24	& 36,876		& 88,421		& 3,167	& 98.4\%		& 97.7\%		& 98.2\%		& 98.5\% \\ \hline
25	& 24,768		& 60,667		& 2,018	& 98.2\%		& 99.6\%		& 98.5\%		& 98.9\% \\ \hline
26	& 16,000		& 39,431		& 1,329	& 98.8\%		& 99.9\%		& 97.6\%		& 98.0\% \\ \hline
27	& 9,216		& 23,764		& 737		& 99.9\%		& 99.7\%		& 97.6\%		& 99.6\% \\ \hline
28	& 3,584		& 10,386		&295		& 98.1\%		& 99.6\%		& 97.7\%		& 98.2\%	\\ \hline
\end{tabular}
}
\end{center}
\label{tab:EvaResult2}
\end{table}

\begin{table*}[ht]
  \caption[EvaResult3]{Results of running symbolic execution with \deepsee\ versus GreenTrie to support constraint solving.}
\begin{center}
\scalebox{.72}{
\begin{tabular}{|c||c|c|c||c|c|c|c|c|c|}
\hline
\multirow{2}{*}{Subjects} & \multicolumn{3}{c||}{SPF with GreenTrie (Large Data Set)} & \multicolumn{6}{c|}{SPF with \deepsee\ (5X5 DNN Trained with Large Data Set)} \\ \cline{2-10} 
 & \# PCs & \# States & Time Cost (s) & \# PCs & Type I Misclassification & Type II Misclassification & \# States & \# Leaf States & Time Cost (s) \\ \hline \hline
 
Red-Black Tree	& 1,329		& 1,330		& 395		& 1,331		& 0		& 2		& 1,332	& 15		& 182		\\ \hline
Dijkstra				& 10,646		& 10,647		& 2,784	& 10,649		& 2		& 3		& 10,650	& 73		& 1,582	\\ \hline
TSP						& 13,212		& 13,213		& 2,418	& 13,215		& 2		& 3		& 13,216	& 50		& 1,189	\\ \hline
Rational				& 744			& 745			& 146		& 748			& 0		& 4		& 749		& 16		& 115		\\ \hline
BinTree					& 3,467		& 3,468		& 907		& 3,579		& 1		& 12		& 3,580	& 37		& 362		\\ \hline
BinomialHeap		& 23,216		& 23,217		& 7,651	& 23,230		& 5		& 14		& 23,231	& 385	& 2,713	\\ \hline
\end{tabular}
}
\end{center}
\label{tab:EvaResult3}
\end{table*}


\subsection{Results and Analysis} \label{result}


For the six subjects that are not used for training DNNs, we first run
symbolic execution with Z3~\cite{Z3Solver}, a state-of-the-art
constraint solving technique, and collect all path conditions with
$11-28$ constraints (as \deepsee\ does not support other path
conditions).  Then, we classify them using a state-of-the-art
constraint solution reuse technique {GreenTrie} and our approach
\deepsee, respectively, and collect data from all three groups of
approaches (including Z3) for evaluation.

For \deepsee, we cross-match two DNN structures and two
training datasets. 
We perform the experiments on the Lonestar 5 cluster at the Texas
Advanced Computing Center (TACC)~\cite{lonestar}. The computing nodes
of Lonestar 5 use Xeon E5-2690 v3 (Haswell) CPU and 64 GB DDR4-2133
memory.  


Table \ref{tab:EvaResult1} shows the results of the experiments. We
report the number of path conditions from each subject (\# PCs) and the 
total time cost of solving them with Z3. For GreenTrie, we calculate the 
total time consumption related to constraint solving including 
pre-processing the PC, visit and retrieving data from the database, 
calling and solving the constraint when a cache miss happens.
We also report the reuse rate of GreenTrie as the percentage of cache
hit to the total invocations. For \deepsee, we report the time cost as
the sum of using deep neural networks to classify PCs
as well as the accuracy of classification results.  In addition,
Table~\ref{tab:EvaResult2} groups the PCs from these six subjects
according to the number of constraints involved ($11-28$), and reports
the results of each individual DNN of \deepsee\ in classifying each
group of PCs. 

{\em RQ1: How accurate is \deepsee\ in path condition classification?}

According to the results in the Table~\ref{tab:EvaResult1}, the
overall accuracy of \deepsee\ is high across different subjects. In
particular, it always achieves over 97.5\% accuracy for
classifying PCs across different subjects. We further look into the
performance of each individual DNN according to
Table~\ref{tab:EvaResult2}, and find that each DNN also achieves over
97.5\% accuracy.



{\em {RQ2:} How efficient is \deepsee\ in path condition
  classification compared to conventional constraint solvers and
  state-of-the-art constraint solution reuse techniques?}  

We observe in Table~\ref{tab:EvaResult1} that \deepsee\ outperforms
{GreenTrie} for all subjects while both \deepsee\ and {GreenTrie} are
faster than conventional constraint solvers as expected. The overall
speedup range of \deepsee\ towards {GreenTrie} is $2.8X$ (with 10
$\times$ 10 DNN structure trained on large database on subject
\texttt{TSP}) to $26.8X$ (with 5 $\times$ 5 DNN structure trained on
small database on subject \texttt{BinomialHeap}), while The speedup
range of \deepsee\ towards {Z3} is $13.6X$ (with 10 $\times$ 10 DNN
structure trained on small database on subject \texttt{BinomialHeap})
to $28.5X$ (with 5 $\times$ 5 DNN structure trained on small database
on subject \texttt{Rational}).  A more intuitive comparison between
the three groups of technique is shown in
Figure~\ref{fig:averCheckingComp}, where we compare the average time
cost of satisfiability checking of a path condition. For {GreenTrie},
we list two different costs of running on small or large database for
reuse, and for \deepsee, we list the average time cost of classifying
the path conditions on the 5 $\times$ 5 DNN trained on the small
dataset. We find that \deepsee\ is significantly faster than Z3 or
GreenTrie, and moreover the cost of \deepsee\ is consistently low
across different subjects.  The performance of {GreenTrie} highly
depends on the reuse rate, as it still needs to call Z3 when there is
no matching of record for reuse. In our evaluation, the overall reuse
rate is not high (even with the large dataset).  However, when the
reuse rate is relatively high (85\% reuse rate for \texttt{TSP} using
the large dataset), its time cost is significantly reduced; however,
it is still outperformed by \deepsee.

{\em {RQ3:} How do the DNN structure and the size of the
  training data impact \deepsee's accuracy and efficiency?}

According to the results, the DNN structure clearly has an impact on
the efficiency of \deepsee.  In particular, \deepsee\ with a larger
DNN model costs more time, since a larger DNN model means the input
data need to go through more layers and neurons. However, the results
have no clear evidence that the structure of DNN can impact on the
accuracy,
as there is no significant difference in accuracy between the two
structures.  On the positive side, it indicates that despite the size
of training dataset, it is possible to use a smaller DNN to achieve
high accuracy while reducing the time cost.

Last but not least, we find that enlarging the dataset does not
necessarily lead to an increase in the reuse rate for
{GreenTrie}. Although its rate is increased from $31\%$ to $85\%$ for
\texttt{TSP}, the increase is only $6\%$ for \texttt{Red-Black
  Tree}. There is no doubt that the size of dataset is an important
factor for the performance of {GreenTrie}, but this shows that
increasing the reuse rate is challenging for constraint reuse
techniques. In contrast, for \deepsee, we are still capable of
training a powerful DNN using a relatively small dataset.

{\em {RQ4:} How well does \deepsee\ support symbolic execution?}

We implemented the Algorithm~\ref{algorithmDeepSee} in Symbolic
Pathfinder (SPF) to use \deepsee\ to support symbolic execution.
Table~\ref{tab:EvaResult3} shows the results of running SPF with
\deepsee\ using 5 $\times$ 5 DNN structure trained with large dataset
compared to running SPF with GreenTrie using the same dataset.  When a
path condition is not supported by \deepsee, we use Z3 to solve
it. For each approach, we report the number of solved/classified PCs,
the number of states, and the total time cost. For SPF with \deepsee\,
we also report the number of misclassification errors as well as the
number of leaf states.


According to the results in the table, 
we find that due to the high accuracy of \deepsee, the number of
each type of misclassification errors are very small compared to the
total PCs.  Since we ignored Type II misclassification errors,
\deepsee\ checked more PCs and explored more states than GreenTrie. In
the meantime, Type I misclassification happened in 4 out of 6
subjects. Despite the overhead introduced in addressing both of the
two types of classification errors, the overall time cost of symbolic
execution with \deepsee\ is still much smaller than symbolic execution
with GreenTrie (e.g., $2.80X$ speedup on \texttt{BinomialHeap}). This
result demonstrates 
that the highly accurate and efficient PC classification in \deepsee\ can
greatly improve the efficiency of symbolic execution.

\subsection{Threats to Validity} \label{threat} 

For external threats to validity, our results may not generalize to
other subjects. Our study was performed on subjects that were used in
previous studies of symbolic execution techniques, and only limited
subjects and versions are suitable for data collection, training and
classification
purposes. 
To mitigate this threat we trained and selected multiple models, and
carefully selected the results that can be potentially generalized,
but it should be noticed that in deep learning studies, it is not
uncommon that a well-trained model cannot be used in a universal
solution to similar but different artifacts. Another threat lies in
the constraint solutions. As expected, the number of satisfiable path
conditions often exceeds the number of unsatisfiable path conditions.
As a result, unlike conventional deep learning techniques, it is very
difficult for us to collect a perfectly balanced dataset for training.
We mitigate this thread by introducing the large training dataset, as
the mutations of subjects contributes a number of unsatisfiable path
conditions which makes the large dataset more balanced than the small
dataset. Based on the evaluation results, our DNNs can achieve more
than 70\% accuracy in classifying unsatisfiable path conditions.

For internal threats to validity, although we have carefully checked
our implementation, it is possible that there are errors we did not
notice. There are also potential threats related to correctness of the
techniques and frameworks we used, including Keras, TensorFlow and
SPF. To mitigate these threats, we treat them as black-box to ensure
that we only made the necessary change to the original {SPF}
implementation. Meanwhile, over-fitting is a common problem when
training a DNN.  To control this threat, we used different techniques
including purifying and shuffling the data, changing the ratio of
training/testing data, using different DNN structures and applying
k-fold validation technique.  The evaluation results show that all
DNNs have a stable and high accuracy on different datasets, and there
is no trace of over-fitting problem with our DNNs. 
\section{Related Work} \label{relateWork}
\subsection{Machine Learning for Constraint Satisfaction
  Problems} \label{CSP} 

Researches have been dedicated to applying machine learning techniques
to constraint satisfaction problems with different models and
techniques including support vector
machines~\cite{Arbelaez:2010:CSC:1916729.1917215}, linear
regression~\cite{DBLP:journals/corr/abs-1111-2249}, decision tree
learning~\cite{Gent:2010:LUL:1860967.1861137,
  Guerri:2004:LTA:3000001.3000101},
clustering~\cite{Kadioglu:2010:IIA:1860967.1861114,
  10.1007/978-3-540-74970-7_41}, k-nearest
neighbors~\cite{o'mahony08usingcase-based}, and so
on~\cite{DBLP:journals/corr/abs-1210-7959}. Xu
et. al~\cite{HXu:TEDLCSP} successfully applied deep learning to
predict the satisfiability of Boolean binary constraint satisfaction
problems with high prediction accuracy. Different from our approach,
this approach uses randomly generated constraint satisfaction problems
as training data and applied a convolutional neural network (CNN) as
the deep learning model, while we take the existing constraint
solutions as a training data set and use a simpler DNN structure.
Moreover, this approach only aims to predict the satisfiability of
Boolean binary constraints, while our approach classify the
satisfiability of path conditions that may have multiple symbolic
variables. Meanwhile, our study is the first to evaluate DNN based
path condition classification in terms of accuracy and efficiency
compared to regular constraint solving and constraint solution reuse
techniques. 

\subsection{Reuse of Constraint Solutions} \label{reuseCS} 

Many techniques have been developed to speed up symbolic execution by
reusing previous constraint solutions. For example,
KLEE~\cite{CadarDE08} optimizes constraint
solving by an approach named counterexample caching. With the cached
constraint solving results, KLEE can quickly check satisfiability of a
path condition if it is a similar query to one of the stored records:
If a path condition has a subset that is already known as unsatifiable,
it is unsatifiable as well. Similarly, if a path condition has an
already known satisfiable superset in the cache, it is satisfiable.

Green~\cite{Visser:2012:GRR:2393596.2393665} applies Redis in-memory
database to maintain the constraint solutions, and uses slicing and
canonizing to path conditions in order to increase the reusing
rate. To further improve Green,
GreenTrie~\cite{Jia:2015} stores constraints and
solutions into L-Trie, which is indexed by an implication partial
order graph of constraints and is able to carry out logical reduction
and logical subset and superset querying for given constraints.
GreenTrie provides more flexibility to conventional Green framework
and expands the number of path conditions that can reuse previous
constraint solutions. Compared to Green and GreenTrie, our approach
reuse the collective knowledge of previous constraint solutions: once
the DNN was trained offline, we do not need to use individual
constraint solutions and can quickly classify the satisfiability of a
path condition as long as it can be transformed into the required form
of matrix.

Unlike techniques that store path conditions and their satisfiability
information, memoized symbolic
execution~\cite{Yang:2012:MSE:2338965.2336771,
  Yang:2013:MTM:2486788.2487001} stores positions and choices taken
during symbolic execution in a trie~\cite{Willard:1984:NTD:859.905} --
an efficient tree-based data structure. When applied to regression
analysis, the trie guided symbolic execution would potentially skip
exploration of portions of program paths, whereas symbolic execution
using our approach would only skip calls to the underlying constraint
solver. Our approach could work together with memoized symbolic
execution to provide a fast classification of path conditions whenever
program paths cannot be skipped by memoized symbolic execution.

Some techniques take advantage of test suites to reduce expensive
constraint solving calls typically in regression testing. For
instance, Makhdoom et al.~\cite{Makhdoom:2014:ISE:2642937.2642961} use
the test suite of a previous program version to check whether a new
path condition is satisfiable or
not. 
Hossain et al.~\cite{6825678} reuse constraint values by comparing the
variables' definitions and uses between program versions. If the
definitions and uses for a certain variable have not changed on a
certain path, constraint values for the variable in the old version
can be reused in the new version.  While these approaches reuse
existing test cases for the purpose of maintaining an effective test
suit for regression testing, our technique is designed to reuse
constraint solving results for the purpose of efficiently classifying
path conditions encountered in symbolic execution of different
programs.



\section{Conclusion and Future Work} \label{conclusionAndFuture}

Symbolic execution is a powerful software engineering analysis
technique, but suffers from the high cost of constraint solving. 
In this paper we introduced \deepsee, a novel approach to solve
constraints based on deep learning, which leverages existing
constraint solutions for training DNNs to classify path conditions for
their satisfiability during symbolic execution.
To the best of our knowledge, this is the first work that results in a
fully functional and applicable solution to use deep learning on
constraint solution reuse for symbolic execution. Our evaluation shows
that \deepsee\ is highly applicable with a high accuracy, is more
efficient than conventional constraint solving and multiple existing
constraint solution reuse frameworks in classifying path conditions
for satisfiability, and can well support overall symbolic execution
task. For future work, we plan to further evaluate our approach on
more real-world artifacts, and compare our solution with other
constraint solution reuse techniques. We also plan to investigate the
use of different DNNs for our approach, e.g., exploring the best DNN
structures for path conditions with different features, and building a
universal DNN for all path conditions.


\section*{Acknowledgments}
This work is partially supported by the National Science Foundation
under Grant Nos. CCF-1464123 and CCF-1659807.

\balance
\bibliographystyle{ACM-Reference-Format}
\bibliography{Bibliography}

\end{document}